\documentclass[twocolumn,showpacs,preprintnumbers,amsmath,amssymb]{revtex4}
\usepackage{graphicx}
\usepackage{dcolumn}
\usepackage{bm}

\begin{document}

\preprint{vajeeston et al }

\title{Pressure-induced structural transitions in MgH${_2}$ }
\author{P. Vajeeston}
 \email{ponniahv@kjemi.uio.no}
\homepage{http://www.folk.uio.no/ponniahv}
\author{P. Ravindran}
\author{A. Kjekshus}
\author{H. Fjellv{\aa}g}
\affiliation{Department of Chemistry,
             University of Oslo, Box 1033, Blindern, N-0315, Oslo, Norway.}
\date{\today}
\begin{abstract}
The stability of MgH$_2$ has been studied up to 20~GPa using density-functional total-energy
calculations. At ambient pressure $\alpha$-MgH${_2}$ takes a TiO$_2$-rutile-type structure.     
$\alpha$-MgH$_2$ is predicted to transform into $\gamma$-MgH${_2}$ at
0.39~GPa. The calculated structural data for $\alpha$- and $\gamma$-MgH${_2}$ 
are in very good agreement with experimental values. At
equilibrium the energy difference between these modifications is
very small, and as a result both phases coexist in a certain volume and pressure field.
Above 3.84~GPa $\gamma$-MgH${_2}$ transforms into
$\beta$-MgH${_2}$; consistent with experimental findings. 
Two further transformations have been identified at still higher pressure: i) 
$\beta$- to $\delta$-MgH${_2}$ at 6.73\,GPa and (ii)
$\delta$- to $\epsilon$-MgH${_2}$ at 10.26~GPa.
\end{abstract}
\pacs{62.50.+P,61.50.Ks,61.66.Fn}
\maketitle
The utilization of high-pressure technology, which can reach the
range of giga Pascal (GPa), has made considerable progress in 
experimental studies of hydrogen storage materials.  It has been
demonstrated for a number of metal$-$hydrogen systems that application of high-pressure 
is an effective tool to produce vacancies
in the host metallic matrix, which in turn leads to various novel properties 
\cite{fukai94}. The Mg-based alloys possess many advantageous
functional properties, such as heat resistance, vibration absorption,
recycling ability, etc. \cite{orimo01}. In recent years,
therefore, much attention has been paid to investigations on specific
material properties of Mg alloys.
Magnesium is an attractive material for
hydrogen-storage applications because of its light weight, low manufacture cost and high
hydrogen-storage capacity (7.66 wt.\,\%). On the other hand, owing to its
high operation temperature (pressure plateau of 1~bar at 525\,K)
and slow absorption kinetics, practical applications of magnesium-based 
alloys have been limited. However, it has recently been established
that improved hydrogen absorption kinetics can be achieved by reduced particle size and/or 
addition of transition metals to magnesium and 
magnesium hydrides \cite{chen95}. $\gamma$-MgH$_2$ often occurs as a
by product in high-pressure syntheses of 
technologically important metal hydrides like Mg$_2$NiH$_4$.
Hence, a complete characterization of $\gamma$-MgH$_2$, in particular, knowledge
about its stability at high pressure is highly desirable.  
Present theoretical investigation assumes importance as the high-pressure 
x-ray diffraction and neutron diffraction studies are unable to identify the
exact position of hydrogen atoms owing to its very-low scattering cross-section. 
This contribution represents the first theoretical report on the high-pressure
properties of MgH$_2$ as evident from {\it ab initio} density functional calculations.

\begin{table*}
\caption{Optimized structural parameters, bulk modulus ($B_{0}$) and its
pressure derivative ($B_{0}^{\prime}$) for MgH$_2$ in the different structural arrangements 
considered in the present study.} 
{
\scriptsize
\begin{ruledtabular}
\begin{tabular}{l l c c c l c c}
Structure & Space group & \multicolumn{3}{c} {Unit cell (\AA)} & Positional parameters & $B_0$ (GPa) & $B_{0}^{\prime}$\\
type      &             & $a$ & $b$ & $c$                            &                        &             &               \\ \hline
TiO$_2$-r$^a$; $\alpha$-MgH$_2$$^b$ & ${P4_2/mnm}$ & 4.4853 & 4.4853 & 2.9993 &  Mg ($2a$): 0,0,0     & 51   & 3.45  \\ 
                 &         & (4.501 & 4.501 & 3.010$^c$) & H ($4f$): .3043,.3043,0; (.304,.304,0$^c$)    & 55$^{d}$; 50$^{e}$  &       \\ 
Mod. CaF$_2$; $\beta$-MgH$_2$$^b$  & ${Pa\bar3}$ & 4.7902 & 4.7902 & 4.7902  & Mg ($4a$): 0,0,0; H ($8c$): .3417,.3417,.3417 & 56    & 3.52   \\   
At 3.84 GPa$^b$                    &             &        &        &         & Mg ($4a$): 0,0,0; H ($8c$): .3429,.3429,.3429 &       &        \\   
$\alpha$-PbO$_2$; $\gamma$-MgH$_2$$^b$ & ${Pbcn}$ & 4.4860 & 5.4024 & 4.8985 &  Mg ($4c$): 0,.3314,1/4 (0,.3313,1/4$^c$)   & 48   & 3.07  \\   
                  &         & (4.501 & 5.4197 & 4.9168$^c$) & H ($8d$): .2717,.1085,.0801 (.2727,.1089,.0794$^c$)  &      &       \\   
At 0.39 GPa$^b$      &         &        &        &             & Mg ($4c$): 0, .3307, 1/4; H ($8d$): .2710,.1073,.0797 &      &       \\   
Ortho; $\delta$-MgH$_2$$^b$ & ${Pbc2_1}$& 4.8604 & 4.6354 & 4.7511 & Mg ($4a$): .0294,.2650,3/4      & 60    & 3.68   \\ 
               &           &        &        &       & H1 ($4a$): .3818,.0976,.8586; H2 ($4a$): .2614,.5584,.0298    &       &        \\ 
At 6.73 GPa$^b$    &           &        &        &       & Mg ($4a$): .0294,.2665,3/4                                    &       &        \\ 
               &           &        &        &       & H1 ($4a$): .3933,.1090,.8475; H2 ($4a$): .2586,.5576,.0305    &       &        \\ 
AlAu$_2$; $\epsilon$-MgH$_2$$^b$  & ${Pnma}$  & 5.2804 & 3.0928 & 5.9903 & Mg ($4c$): 1/4,3/4,.6033; H1 ($4c$): .3610,1/4,.4250    & 65    & 3.72   \\   
               &           &        &        &       & H2 ($4c$): .4794,1/4,.8345    &       &        \\   
At 10.2 GPa$^b$    &           &        &        &       & Mg ($4c$): 1/4,3/4,.6075; H1 ($4c$): .3602,1/4,.4260       &       &        \\   
               &           &        &        &       & H2 ($4c$): .4765,1/4,.8300    &       &        \\   
InNi$_2$$^b$  & ${P6_3}$ & 3.2008 & 3.2008 & 5.9870 & Mg ($2b$): 1/3,2/3,1/4; H1 ($2a$): 0,0,0   & 59    &  3.42  \\   
               &           &        &        &        & H2 ($2b$): 1/3,2/3,3/4   &       &        \\   
Ag$_2$Te$^b$  & ${P2_1/c}$ & 4.8825 & 4.6875 & 5.0221 & Mg ($4e$): .2630,.4788,.2117    & 44    & 3.31   \\ 
              &            &        & ($\beta$ = 99.25 )  &        & H1 ($4e$): .0718,.2039,.3818; H2 ($4e$): .4309,.7430,.5021    &       &        \\ 
CaF$_2$$^b$     & ${Fm}{\bar3m}$ & 4.7296 & 4.7296 & 4.7296 & Mg ($4a$): 0,0,0; H ($8c$): 1/4,1/4,1/4 & 59 & 3.52   \\ 
TiO$_2$-a$^{b,f}$  & ${I4_1/amd}$ & 3.7780 & 3.7780 & 4.7108 & Mg ($4a$): 0,0,0; H ($8e$): 0, 0, .2101   & 45    & 3.19   \\   
AuSn$_2$$^b$  & ${Pbca}$  & 9.3738 & 4.8259 & 4.7798 & Mg ($8c$): .8823,.0271,.2790; H1 ($8c$): .7970,.3765,.1651 & 58    & 3.45   \\   
               &           &        &        &       & H2 ($8c$): .9738,.7433,.5207    &       &        \\   
CaCl$_2$$^b$  & ${Pnnm}$ & 4.4810 & 4.4657 & 3.0064 & Mg ($2a$): 0,0,0; H ($4g$): .3076,.2991,0 & 58  & 3.45   \\   
\end{tabular}
\end{ruledtabular}
}
\label{table:str}
\footnotetext{$^a$ r = rutile. $^b$ To elucidate variations of atomic positions with pressure, parameter values are given at equilibrium volume           
as well as at transition pressures. $^c$Experimental value \cite{bortz99}. $^{d,e}$Theoretical value \cite{prommer94}, \cite{yu88}.                   
$^f$ a = anatase.} 
\end{table*}

\begin{figure}
\includegraphics[scale=0.425]{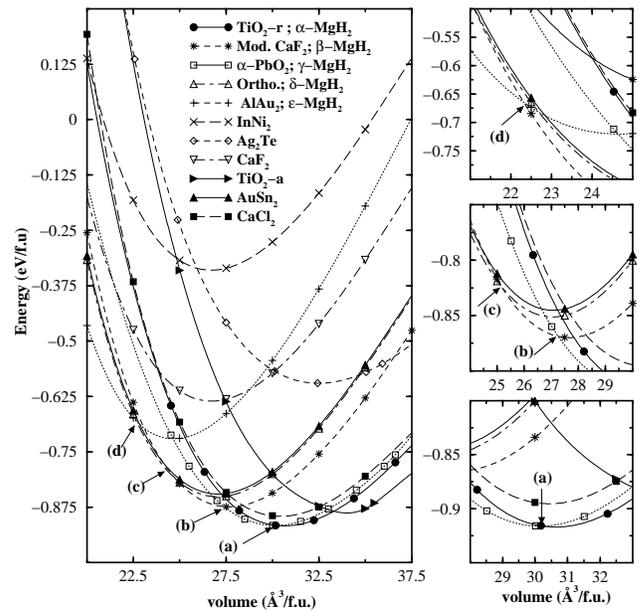}
\caption{Calculated unit cell volume vs. total energy relations for MgH$_2$ 
in actual and possible structural arrangements as obtained from VASP. 
Magnified version of the corresponding transition points are shown on right hand side of the figure.}
\label{fig:ene}
\end{figure}

\par
$\alpha$-MgH$_2$ crystallizes with TiO$_2$-r-type (r = rutile) structure 
at ambient pressure and low temperature~\cite{bastide80,zachariasen63}.
At higher temperatures and pressures tetragonal $\alpha$-MgH$_2$ transforms
into orthorhombic $\gamma$-MgH$_2$.
Recently Bortz {\em et al.}~\cite{bortz99} solved the crystal
structure of $\gamma$-MgH$_2$ ($\alpha$-PbO$_2$ type) on the basis of powder
neutron diffraction data collected at 2~GPa. The $\alpha$- to $\gamma$-MgH$_2$ 
transition pressure is not yet known.  Owing to the
difficulties challenged in establishing hydrogen position in a metal matrix
by x-ray diffraction high-pressure information on MgH$_2$ is scarce. 
The aim of the present investigation is to remedy this situation by examining 
MgH$_2$ theoretically at high pressure in eleven closely related structural configurations.

\par
We used the {\it ab initio} generalized gradient approximation (GGA)
\cite{perdew} to obtain accurate exchange and correlation energies for a particular
structural arrangement. The structures are fully relaxed for all
the volumes considered in the present calculations using force
as well as stress minimization.  Experimentally established 
structural data were used as input for the calculations when available.  The full-potential linear muffin-tin
orbital (FP-LMTO)\cite{wills} and the projected augmented wave (PAW) implementation of Vienna {\it ab initio} simulation
package (VASP) \cite{vasp1} were used for the total-energy
calculations to establish phase stability and transition pressures.  
Both methods yielded nearly the same result,
e.g., the transition pressure from $\alpha$- to $\gamma$-MgH$_2$ came out as 
0.385 and 0.387~GPa from FP-LMTO and VASP, respectively. 
Similarly almost identical values were obtained  for  
ground-state properties like bulk modulus and equilibrium volume.
Hence, for rest of the calculations we used only the VASP code
because of its computational efficiency.  In order to avoid 
ambiguities regarding the free-energy results we have always used the same energy cutoff
and a similar {\bf k}-grid density for convergence. In all 
calculations 500 {\bf k} points in the whole Brillouin
zone were used for $\alpha$-MgH$_2$ and a similar density of {\bf k} points 
for the other structural arrangements. The PAW pseudo
potentials \cite{paw} were used for all our VASP calculations. A criterion of at least 0.01 meV/atom was
placed on the self-consistent convergence of the total energy, and the
calculations reported here used a plane wave cutoff of 400\,eV.
\begin{figure}
\includegraphics[scale=0.525]{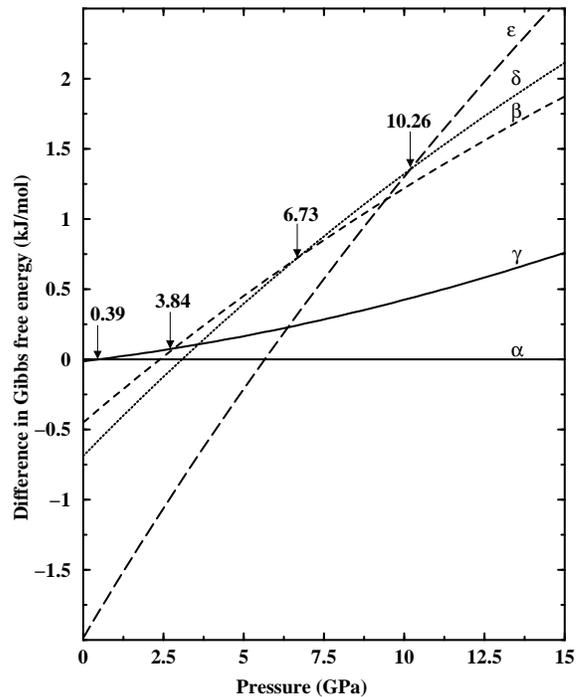}
\caption{The stabilities of various known and hypothetical MgH$_2$ phases
relative to $\alpha$-MgH$_2$ as a function of pressure. The transition pressures
are marked as a arrow at the corresponding transition points.} 
\label{fig:gibbsen}
\end{figure}
\par
In addition to the experimentally identified $\alpha$- and $\gamma$-modifications of MgH$_2$,
we have carried out calculations for several other possible \cite{lowther99} 
types of structural arrangements 
for MgH$_2$ (see Table~\ref{table:str} and Figs.~\ref{fig:ene} and ~\ref{fig:gibbsen} ). 
The calculated total energy vs. volume relation for all these alternatives 
are shown in the Fig.~\ref{fig:ene}. 
In order to get a clear picture about the structural transition points, in Fig.~\ref{fig:gibbsen} we have displayed 
the Gibbs free energy difference between the pertinent crystallographic structures of MgH$_2$
with reference to $\alpha$-MgH$_2$ as a function of pressure.  
The equilibrium volumes,
(30.64 and 30.14\,\AA${^3}$/f.u. for $\alpha$- and 
$\gamma$-MgH$_2$, respectively) are within 1\,\% of the experimental values indicating that the
theoretical calculations are reliable. The calculated positional
parameters (Table~\ref{table:str}) are also in
excellent agreement with the experimental data. The calculated values
for the bulk modulus ($B_0$, see Table~\ref{table:str}) vary between 44 to 65~GPa for the various
structural arrangements. Among the eleven possibilities considered the Ag$_2$Te-type leads to the smallest
$B_0$ value and the AlAu$_2$-type to the highest (Table~\ref{table:str}).

\begin{figure}
\includegraphics[scale=0.5  ]{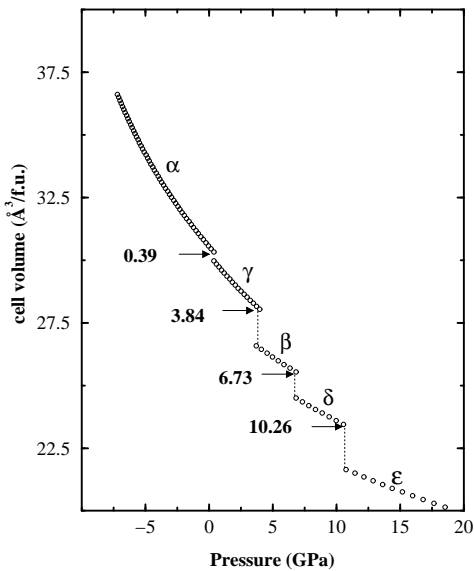}
\caption{Pressure vs. volume relation for MgH$_2$. Pressure stability regions for the 
different modifications (see Table~\ref{table:str} and text) are indicated.} 
\label{fig:pv}
\end{figure}

\par
The calculated transition pressure for the $\alpha$- to $\gamma$-MgH$_2$ 
conversion is 0.387~GPa (Fig.~\ref{fig:pv}) and as the free
energy of the two modification is nearly the same at the equilibrium volume (Fig.~\ref{fig:ene}),
it is only natural that 
these phases coexist in a certain volume range \cite{bortz99}. Structurally $\alpha$- and
$\gamma$-MgH$_2$ are also closely related, both comprising Mg in an octahedral coordination 
of 6 H which in turn are linked by edge sharing in  
one direction and by corner sharing in the two other directions.  The
chains are linear in tetragonal $\alpha$-MgH$_2$ and run
along the four-fold axis of its TiO$_2$-r-type structure, while they are
zig-zag shaped in $\gamma$-MgH$_2$ and run along a two-fold
screw axis of its orthorhombic $\alpha$-PbO$_2$-type structure. The
octahedra in $\gamma$-MgH$_2$ are strongly distorted. The pressure
induced $\alpha$-to-$\gamma$ transition involves reconstructive (viz. bonds are 
broken and re-established) rearrangements of the cation and anion sublattices.
The occurrence of a similar phase transition in PbO$_2$ suggests that thermal activation
or enhanced shear is of importance \cite{haines96}.

\begin{figure}
\includegraphics[scale=0.425]{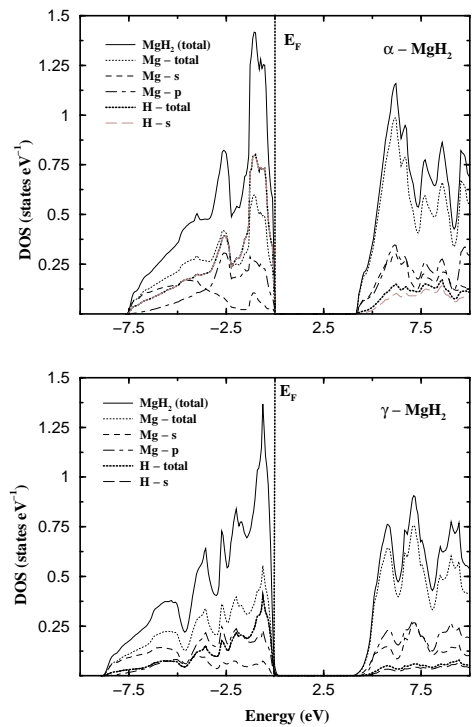}
\caption{The electronic total and partial density of states 
for $\alpha$- and $\gamma$-MgH$_2$.}
\label{fig:dos}
\end{figure}
\par
The subsequent phase transition from $\gamma$- to 
$\beta$-MgH$_2$ occur at 3.84~GPa.  Bortz {\em et al.}~\cite{bortz99} found no
evidence for the formation of such a $\beta$ modification up to 2~GPa, whereas
Bastide {\it et al.}~\cite{bastide80} found that at higher pressure
(4~GPa) and temperature (923~K) there occurs a new phase (viz. $\beta$ 
in accordance with our findings).
However, the H positions are not yet determined experimentally and hence our optimized structural
parameters for $\beta$-MgH$_2$ should be of particular interest. The
volume shrinkage at the transition point is 1.45 {\AA}$^3$/f.u. On further increase
of the pressure to around 6.7~GPa, $\beta$-MgH$_2$ is predicted to transform into
$\delta$-MgH$_2$ (orthorhombic $Pbc2_1$) with a volume shrinkage of 
1.1~{\AA}$^3$/f.u. In the pressure range 6.7$-$10.2~GPa the structural arrangements 
$\beta$-MgH$_2$, AuSn$_2$ and $\delta$-MgH$_2$
are seen to lie within a narrow 
energy range of some 10 meV (Fig.~\ref{fig:ene}). This closeness in energy suggests that the
relative appearance of these modification will be quite sensitive to, and easily
affected by, other external factors like temperature and remnant lattice
stresses.  A transformation to $\epsilon$-MgH$_2$ (AlAu$_2$-type structure also called cotunnite-type) 
is clearly evident from Fig.~\ref{fig:ene} with a 
volume change of 1.7~{\AA}$^3$/f.u. at the transition point (see also Figs.~\ref{fig:gibbsen} and ~\ref{fig:pv}).
It is interesting to note that similar structural transition sequences are reported for 
transition-metal oxides like HfO$_2$~\cite{lowther99}. Recently, the cotunnite-type structure of
TiO$_2$ (synthesized at pressure above 60 GPa and at high temperatures) has been shown \cite{dubrovinsky01} to
exhibit an extremely high bulk modulus (431 GPa) and hardness (38 GPa). 
Our present study predicts that, it is possible to stabilize MgH$_2$ in the AlAu$_2$-type structure as 
a soft material ($B_0$ = 65 GPa) above 10.26~GPa. Hence, the present 
prediction of stabilization of a contunnite-type structure for MgH$_2$ at high pressure should be 
of considerable interest.

\par
The calculated unit-cell volume difference between the involved phases at their equilibrium
volume relative to the $\alpha$-phase is 0.497 (for $\gamma$), 2.94 (for $\beta$),
3.60 (for $\delta$), and 5.96 (for $\epsilon$) \,\AA${^3}$/f.u., which is
approximately 1.6, 9.5, 11.8, and 19.5\,\%, respectively, smaller than the equilibrium
volume of the $\alpha$-phase. The energy difference between $\alpha$-phase
and the $\gamma$, $\beta$, $\delta$, and $\epsilon$ phases at their equilibrium volume 
is 0.81, 43, 66, and 197 meV/f.u., respectively, which
is considerable smaller than for similar types of phase transitions in 
transition metal dioxides \cite{lowther99}. 
The known stabilization of the high-pressure phase of TiO$_2$ by using high-pressure$-$high-temperature
synthesis \cite{dubrovinsky01} indicates that it may be possible to stabilize the high-pressure
phases of MgH$_2$ due to their small energy differences and (partly) reconstructive
phase transitions.  Such stabilization of MgH$_2$ high-pressure phases at room temperature
would reduce the volume considerably, and 
in the extreme case the
expected volume reduction would be ca. 19.5\,\% for $\epsilon$-MgH$_2$ compared with $\alpha$-MgH$_2$. 
This would imply an increased volumetric storage capacity of hydrogen for MgH$_2$.
Of interest would be to explore the possibility of stabilizing the high-pressure phases by
chemical means. At the same time this may possibly open up for improved kinetics with
respect to reversible hydrogen absorption/desorption. Hence, the observation of high-pressure
phases in MgH$_2$ may have technological significance.

\par
The calculated energy gap (from the top of the valence band (VB) to bottom of the
conduction band) is 4.2 and 4.3\,eV for $\alpha$- and $\gamma$-MgH$_2$,
respectively. According to Fig.~\ref{fig:dos} the
width of VB is 7.5 and 8.3\,eV in
$\alpha$- and $\gamma$-MgH$_2$, respectively. The
increased width of VB in $\gamma$-MgH$_2$ is due to the reduction
in the Mg$-$H separation (1.93~{\AA} in $\alpha$-MgH$_2$ vs. 1.91~{\AA} in $\gamma$-MgH$_2$). 
An experimental UV absorption study \cite{prommer94} of $\alpha$-MgH$_2$ gave an absorption edge at
5.16\,eV, whereas earlier theoretical electronic structure investigation \cite{yu88,prommer94} 
reported a band gap around 3.4\,eV.
Hence, our calculated energy
gap value is in better agreement with the experimentally measured value
than the earlier values.
In fact, the present underestimation of the band gap by about
17\,\% is typical for the accuracy obtained by first principle methods
for semiconductors and insulators. Such distinctions
probably originates from the use of GGA-based exchange
correlation functionals.

\par
In conclusion, the calculated structural parameters for $\alpha$- and $\gamma$-MgH$_2$
are in excellent agreement with the experimental findings. 
We found that the ground state of 
$\alpha$-MgH$_2$ becomes unstable at higher pressure and the calculated transition pressures for
$\alpha$-to-$\gamma$- and $\gamma$-to-$\beta$-MgH$_2$ are 
found to be in very good agreement with the experimental findings.
We predict that further compression of
$\beta$-MgH$_2$ will lead to a phase transition to $\delta$-MgH$_2$ (orthorhombic; ${Pbc2_1}$) at 6.73\,GPa
and that the AlAu$_2$-type structure of a hypothetical $\epsilon$ modification stabilizes above 10.26\,GPa.
Our total energy study suggested that the formation of the high-pressure $\epsilon$-phase could also be
possible at the ambient pressure/temperature by choosing an appropriate preparation method. This may 
reduce the volume requirement by ca. 19.5\,\%  compared with $\alpha$-MgH$_2$.
In agreement with experiments, $\alpha$-MgH$_2$ turns out to be an insulator, and the 
theoretical treatment shows that the high-pressure modifications also should exhibit insulating behavior. 
\par  
The authors greatfully acknowledge the Research Council of Norway for
financial support and for the computer time at the Norwegian supercomputer
facilities and R. Vidya for critical reading of this manuscript.

\end{document}